\documentstyle[12pt]{article}

\begin{document}
\mbox{ }
\rightline{UCT-TP-246/97}\\
\rightline{December 1997}\\
\vspace{3.5cm}
\begin{center}
{{\Large \bf The Strange-Quark Mass from QCD}\\[.3cm]
{\Large \bf Sum Rules in the Pseudoscalar Channel}} \\
\vspace{.5cm}
{\bf C. A. Dominguez$^{(a)}$, L. Pirovano$^{(a)}$, and K. 
Schilcher$^{(b)}$}\\[.5cm]
$^{(a)}$Institute of Theoretical Physics and Astrophysics\\
University of Cape Town, Rondebosch 7700, South Africa\\[.5cm]
$^{(b)}$Institut f\"{u}r Physik, Johannes Gutenberg-Universit\"{a}t\\
Staudingerweg 7, D-55099 Mainz, Germany
\end{center}
\vspace{.5cm}
\begin{abstract}
\noindent
QCD Laplace transform sum rules, involving the axial-vector current 
divergences, are used in order to determine the strange quark mass. 
The two-point function is known in QCD up to four
loops in perturbation theory, and up to dimension-six
in the non-perturbative sector. The hadronic spectral
function is reconstructed using threshold normalization
from chiral symmetry, together with experimental data for
the two radial excitations of the kaon. The result for the 
running strange quark mass, in the $\overline{MS}$ scheme
at a scale of 1 $\mbox{GeV}^{2}$  is:
${\bar m}_{s}(1 GeV^{2}) = 155 \pm 25 \mbox{MeV}$.
\end{abstract}
\newpage
\setlength{\baselineskip}{1.5\baselineskip}
\noindent
In spite of many attempts \cite{NAR}-\cite{BIJ} to improve the accuracy of
QCD sum rule determinations of the strange quark mass, present uncertainties
still remain uncomfortably large. This is a serious limiting factor affecting
areas such as  kaon-physics and CP violation, which depend strongly on
$m_{s}$. One of the determinations believed to be among the most accurate
was the one that combined the current algebra ratio \cite{GL} 
\begin{equation}
\frac{m_{s}}{m_{u}+m_{d}} = 12.6 \pm 0.5 \; ,
\end{equation}
together with a QCD sum rule determination of ($m_{u}+m_{d}$). The latter
was discussed some years ago \cite{CAD1} in the framework of QCD Finite
Energy Sum Rules (FESR) in the pseudoscalar channel,
at the two-loop level in perturbative QCD,
and including non-perturbative condensates up to dimension-six, with
the result
\begin{equation}
(m_{u}+m_{d}) \; \mbox{(1 GeV)} = 15.5 \pm \mbox{2.0 MeV} \; .
\end{equation}
This result, together with Eq.(1), implies
\begin{equation}
m_{s} \; \mbox{(1 GeV)} = 195 \pm \mbox{28 MeV} .
\end{equation}
A recent re-analysis \cite{BIJ} of the same QCD sum rules, 
but including the next  (3-loop) order in perturbation theory, 
and a somewhat different hadronic spectral function, obtains
\begin{equation}
(m_{u}+m_{d}) \; \mbox{(1 GeV)} = 12.0 \pm \mbox{2.5 MeV} \; ,
\end{equation}
which, using Eq.(1), leads to
\begin{equation}
m_{s} \; \mbox{(1 GeV)} = 151 \pm \mbox{32 MeV} \; .
\end{equation}
The same raw data for resonance masses
and widths, plus the same threshold normalization from chiral
perturbation theory, has been used in both analyses
\cite{CAD1} and \cite{BIJ}. The difference in the results 
cannot be accounted for by the inclusion (or not) of the 3-loop perturbative
QCD contribution, which amounts to a reduction of the 2-loop result, Eq.(2),
of only a few percent. Instead, this difference stems mainly
from the different reconstructions of the spectral function from
the resonance data; with a more elaborate functional form  being adopted
in \cite{BIJ}. 
This reveals a type of {\it systematic} uncertainty of the QCD sum rule
method, that will cease to be an uncertainty only after the pseudoscalar
hadronic spectral function is measured {\it directly}
and accurately (e.g. from tau-lepton decays). In the absence of such
direct data, it is not possible to exclude the presence of a background
(constructive or destructive) beyond that implicit in the chiral 
normalization. Also, this overall normalization could assume a 
non-standard value as advocated in \cite{STERN}.
Hence, at present, both results for $m_{s}$, Eqs. (3) and (5), are equally 
acceptable, and taken together provide a measure of underlying 
systematic uncertainties.\\[.3cm]
A second alternative is based on QCD sum rules involving the correlator
of the strange vector current divergences.
In this case, the availability of experimental
data on $K-\pi$ phase shifts \cite{EXP} allows, in principle, for a
reconstruction of the hadronic spectral function
in this channel, from threshold up to
$s \; \simeq \; \mbox{7 GeV}^{2}$. In both \cite{CH1} and \cite{J1},
the functional form chosen for this reconstruction consisted of a
superposition of two Breit-Wigner resonances, corresponding to the
$K_{0}^{*}$ (1430) and the $K_{0}^{*}$ (1950) \cite{PDG},
normalized at threshold according to 
conventional chiral-symmetry. It was argued in \cite{CH1} and 
\cite{J1} that the non-resonant background implicit in this threshold
normalization was  important to achieve a good fit to the
$K-\pi$ phase shifts. In \cite{CH1} and \cite{J1}, the relevant correlator  
was calculated in perturbative QCD at the 3-loop level, with mass
corrections up to the quartic order, and including non-perturbative
quark and gluon vacuum condensates up to dimension four (the $d=6$
condensates are numerically unimportant and can be 
safely neglected \cite{J1}). Using Laplace transform sum rules, the results
for the strange quark mass thus obtained were
\begin{eqnarray}
m_{s} \; \mbox{(1 GeV)} \; = \; 
\begin{array}{lcl}
171 \pm 15 \; \mbox{MeV} \; \; \; (\cite{CH1})\\[.3cm]
178 \pm 18 \; \mbox{MeV} \; \; \; (\cite{J1})
\end{array}
\end{eqnarray}
The errors reflect uncertainties in the experimental data, in the value
of $\Lambda_{\mbox{QCD}}$ ($\Lambda_{\mbox{QCD}} \simeq$ 200 - 500 MeV),
in the
continuum threshold $s_{0}$ ($s_{0} \simeq \mbox{6 - 7 \mbox{GeV}}^{2}$), 
and in the values of the vacuum condensates.\\ [.3cm]
As mentioned earlier, an example of a potential systematic error 
affecting these results would be the presence of a background, beyond
the one implicit in the chiral-symmetry normalization of the 
hadronic spectral function at threshold. Obviously, this is not included
in (6). A reanalysis of this QCD sum rule determination  of $m_{s}$
\cite{IT1} has uncovered this uncertainty. In fact, it is claimed                                                        
in \cite{IT1} that by using the Omnes representation to relate the
spectral function to the $K-\pi$ phase shifts, it is necessary to
include a background interfering {\it destructively} with the resonances. As
a result, the hadronic spectral function is considerably smaller than
that used in \cite{CH1}-\cite{J1}. This in turn implies smaller values
of $m_{s}$, viz.
\begin{equation}
m_{s} \; \mbox{(1 GeV)} = 140 \pm 20 \mbox{MeV} \; .
\end{equation}

Still another source of systematic uncertainty, this time of a theoretical
nature, has been unveiled in \cite{CH2}. This has to do with the 
following two possibilities: (a) expand the QCD correlator in inverse
logarithms of the momentum transfer $Q^{2}$, or (b) expand only in
terms of powers of the strong coupling $\alpha_{s}$. Similarly, after
Laplace transforming the correlator one faces the same problem, except
that the momentum transfer is being replaced by the Laplace variable
$M^{2}$.
It has been argued in \cite{CH2} that it makes
more sense to make full use of the  perturbative expansions of the quark
mass and coupling (known to 4-loop order), and hence not to expand them.
Numerically, it turns out that the non-expanded expression is
far more stable than the truncated one, when moving from one order in 
perturbation theory to the next. This fact lends strong support to the
non-expanded alternative. In addition, as shown in \cite{CH2},
logarithmic truncation can lead to sizable overestimates of radiative
corrections. This in turn implies an underestimate of the quark mass.
In fact, after using untruncated expressions, together with the same
hadronic spectral function parametrization as in \cite{CH1}-\cite{J1},
the authors of \cite{CH2} find
\begin{equation}
 m_{s} \; \mbox{(1 GeV)} = 203 \pm 20 \mbox{MeV} \; ,
 \end{equation}
to be compared with the results (6) obtained from truncated expressions.
Until the questions of truncation, and of the correct form of the
hadronic spectral function become satisfactorily settled,
one should take the value (8) together with (6), and include (7) as well.
This would give $m_{s} \; \mbox{(1 GeV)} = 170 \pm 50 \mbox{MeV}$,
a rather inaccurate, albeit realistic result.\\[.3cm]
In this note we discuss a direct determination of $m_{s}$ using QCD Laplace 
sum rules in the pseudoscalar channel, i.e. involving the correlator 
\begin{eqnarray}
 \psi_{5} (q^2) = i \; \int \; d^4 \; x \; e^{i q x} \; \;
  <0|T(\partial^{\mu} \; A_{\mu}(x) \; \partial^{\nu}
   \; A_{\nu}^{\dagger}(0))|0> \; ,
\end{eqnarray}
where $A_{\mu}(x) = :\bar{s}(x)  \gamma_{\mu}  \gamma_{5} u(x):$, and
$\partial^{\mu} \; A_{\mu}(x) = m_{s} \; :\bar{s}(x)  i  \gamma_{5}
\;  u(x): \;$.
The QCD expression of this two-point function is known \cite{CH1},\cite{J1},
\cite{CH2} at the four-loop level in perturbative QCD, and up to dimension
six in the non-perturbative sector. Also, the old problem of mass 
singularities has been satisfactorily solved in \cite{CH1},\cite{J1}. As 
a result of this, quark mass corrections are also known up to quartic
order. Notice that the QCD result for the correlator (9) is trivially
obtained from that involving the vector current divergences; hence, they
both look quite similar. The QCD expression of the
Laplace transform of Eq.(9), i.e. 
\begin{equation}
\psi_{5}^{''}(M^{2}) = \hat{L} \left[ \psi_{5}^{''}(Q^{2}) \right] = 
\int_{0}^{\infty} \;
e^{-s/M^{2}} \; \frac{1}{\pi} \; \mbox{Im} \;\; \psi_{5}(s) \; ds \; ,
\end{equation}
is given by
\begin{equation}
\psi_{5}^{''}(M^{2})|_{QCD} = \left[ \bar{m}_{s} (M^{2}) \right]^{2} \;
M^{4} \left[ \psi_{5(0)}^{''} 
(M^{2}) + \frac{\psi_{5(2)}^{''}(M^{2})}{M^{2}} 
+ \frac{\psi_{5(4)}^{''}(M^{2})}{M^{4}}
+ \frac{\psi_{5(6)}^{''}(M^{2})}{M^{6}} + \cdots \; \right] \; ,
\end{equation}
where
\begin{eqnarray*}
\psi_{5(0)}^{''}(M^{2}) \equiv \hat{L} \; [\psi_{5(0)}^{''}(Q^{2})] =
\frac{3}{8 \pi^{2}}
\Biggl\{ 1 + \frac{\bar{\alpha}_{s}(M^{2})}{\pi} \left( \frac{11}{3}  
+ 2 \gamma_{E} \right) + \left( \frac{\bar{\alpha}_{s}(M^{2})}{\pi} 
\right)^{2} \left( \frac{5071}{144} \right.
\end{eqnarray*}
\begin{eqnarray*}
\left. - \frac{35}{2} \; \zeta (3) + \frac{17}{4} \; \gamma_{E}^{2} 
 + \frac{139}{6} \; \gamma_{E} - \frac{17}{24} \; \pi^{2} \right) 
+ \left( \frac{\bar{\alpha}_{s}(M^{2})}{\pi} \right)^{3} \left( - 
\frac{4781}{9} + \frac{1}{6} \; a_{1} \right.
\end{eqnarray*}
\vspace{.2cm}
\begin{equation}
\Biggl. \left. - \frac{475}{4} \; \zeta (3) \; \gamma_{E} + \frac{823}{6} \;
 \zeta (3) + 
\frac{221}{24} \; \gamma_{E}^{3} + \frac{695}{8} \; \gamma_{E}^{2}
- \frac{221}{48} \; \gamma_{E} \; \pi^{2} + \frac{2720}{9} \; \gamma_{E} -
\frac{695}{48} \; \pi^{2} \right) \Biggr\} \; ,
\end{equation}
\newpage
\begin{equation}
 \psi_{5(2)}^{''}(M^{2}) \equiv \hat{L} \left[ \psi_{5(2)}^{''} 
 (Q^{2}) \right] = -
\frac{3}{4 \pi^{2}} \; \left[ \bar{m}_{s} (M^{2}) \right]^{2} \left[ 1 + 
\frac{\bar{\alpha}_{s}(M^{2})}{\pi} \left( \frac{16}{3} + 4 \gamma_{E} 
\right) \right] \; ,
\end{equation} 
\vspace{.3cm}
\begin{eqnarray*}
\psi_{5(4)}^{''} (M^{2}) \equiv \hat{L} \left[ \psi_{5(4)}^{''} (Q^{2}) \right] = 
\frac{1}{8} \;
< \frac{\alpha_{s}}{\pi} \; G^{2} > + \frac{1}{2} \; < m_{s} \; \bar{s} s > 
\;
\left[ 1 + \frac{\bar{\alpha}_{s}}{\pi} \left( \frac{11}{3} + 2 \gamma_{E} 
\right) \right]
\end{eqnarray*}
\begin{eqnarray*}
- < m_{s} \; \bar{u} u > \; \left[ 1 + \frac{\bar{\alpha}_{s}}{\pi} \left(
 \frac{14}{3} + 2 \gamma_{E} \right) \right]
\end{eqnarray*}
\begin{equation}
+ \frac{3}{28 \pi^{2}} \; m_{s}^{4} \left[ - \frac{233}{36} - \frac{15}{2} \;
\gamma_{E} + 2 \frac{\bar{\alpha}_{s}}{\pi} \left( \frac{37}{9} + 
2 \gamma_{E} 
\right)  \left( \frac{\pi}{\bar{\alpha}_{s}} - \frac{53}{24} \right) 
\right] \; ,
\end{equation}

and where $\gamma_{E}$ is Euler's constant,  $\zeta(n)$ is Riemann's zeta
function, $a_{1} = 2795.0778$, all numerical coefficients refer to
three flavours and three colours,
and  we have neglected the up-quark mass everywhere. Given the
uncertainties of the method, plus the size of systematic errors, it is
not justified to keep $m_{u}$ different from zero.
The four-loop expressions for the strong running coupling and quark mass
are given by\\[.3cm]
\begin{equation}
\frac{\bar{\alpha}_{s}(M^{2})}{\pi} = \frac{4}{9} \; \frac{1}{L} -
\frac{256}{729} \; \frac{LL}{L^{2}} + \left[ 6794 - 16384 \; (LL - LL^{2})
\right] \; \frac{1}{59049} \; \frac{1}{L^{3}},
\end{equation}
\begin{eqnarray*}
\bar{m}_{s}(M^{2}) = \frac{\hat{m}_{s}}{(\frac{1}{2} L)^{4/9}} \;
\Biggl\{ 1 + (290 - 256 LL) \; \frac{1}{729} \; \frac{1}{L} +
 \left[ \frac{550435}{1062882} - \frac{80}{729} \; \zeta (3) \right.
\end{eqnarray*}
\begin{eqnarray*}
\left. - \; (388736LL - 106496 \; LL^{2})
\; \frac{1}{531441} \right] \; \frac{1}{L^{2}} + \left[
- \frac{126940037}{1162261467} - \frac{256}{177147} \; \beta_{4} \right.
\end{eqnarray*}
\begin{eqnarray*}
+ \; \frac{128}{19683} \; \gamma_{4} + \frac{7520}{531441} \; \zeta (3) 
+ \left( - \frac{611418176}{387420489} + \frac{112640}{531441} \; \zeta (3)
\right) \; LL
\end{eqnarray*}
\vspace{.3cm}
\begin{equation}
\left. \Biggl. + \; \frac{335011840}{387420489} \; LL^{2}
- \frac{149946368}{1162261467} \; LL^{3} \right] \; \frac{1}{L^{3}}
\Biggr\} ,
\end{equation}

where $L = log (M^{2}/\Lambda_{QCD}^{2})$, $LL = log L$,
and  \cite{beta4}

\begin{equation}
\beta_{4} = - \frac{281198}{4608} - \frac{890}{32} \zeta (3),
\end{equation}

with  $\gamma_{4} = 88.5258$ (see \cite{gamma4}). In  line with the
discussion after Eq. (7), and following \cite{CH2}, we shall not expand
the above QCD expressions in inverse powers of $L$, but rather
substitute the numerical values of $\alpha_{s}(M^{2})$ 
and $\bar{m}_{s}(M^{2})$ as determined from Eqs.(15)-(16)
for a given value of $\Lambda_{QCD}$. The dimension-six
non-perturbative term has been omitted as it is of no numerical importance.

The hadronic spectral function associated with the correlator (9)
is very different from that of the vector divergences. There
is, at present, preliminary information from tau-decays \cite{ALEPH} in a 
kinematical range restricted by the tau-mass. We  reconstruct the
spectral function, including in addition to the kaon-pole its radial
excitations K(1460) and K(1830), normalized at threshold according to
conventional chiral symmetry. 
In addition, we  incorporate the resonant sub-channel  $K^{*}(892)-\pi$, 
which is of numerical importance given the narrow width of the $K^{*}(892)$
(the sub-channel $\rho (770)-K$ is numerically negligible). This chiral
symmetry normalization is of the form
\vspace{.3cm}
\begin{equation}
\frac{1}{\pi} \; \mbox{Im} \; \psi_{5}(s)|_{K \pi \pi} =
\frac{M_{K}^{2}}{2f_{\pi}^{2}} \; \frac{3}{2^{8} \pi^{4}} \;
\frac{I(s)}{s(M_{K}^{2} - s)} \; \theta (s - M_{K}^{2}) \;,
\end{equation}
where
\begin{eqnarray*}
I(s) = \int_{M_{K}^{2}}^{s} \; \frac{du}{u} \; (u - M_{K}^{2}) \; (s - u) \;
\Biggl\{ (M_{K}^{2} - s) \left[ u - \frac{(s+M_{K}^{2})}{2} \right] \Biggr.
\end{eqnarray*}
\begin{equation}
\Biggl. - \frac{1}{8u} \; (u^{2} - M_{K}^{4}) \; (s - u) + \frac{3}{4} \;
(u - M_{K}^{2})^{2} |F_{K^{*}} (u)|^{2} \Biggr\} \; ,
\end{equation}
and
\begin{equation}
|F_{K^{*}} (u)|^{2} = \frac{ \left[ M_{K^{*}}^{2} - M_{K}^{2} \right]^{2} +
M_{K^{*}}^{2} \; \Gamma_{K^{*}}^{2}}
{ (M_{K^{*}}^{2} - u)^{2} + M_{K^{*}}^{2} \; \Gamma_{K^{*}}^{2}} \; . 
\end{equation}

The pion mass has been neglected above, in line with the approximation
$m_{u} = 0$ made in the QCD sector, and in our normalization
$f_{\pi} \simeq 93 MeV$. The complete hadronic spectral function is then

\begin{eqnarray*}
\frac{1}{\pi} \; \mbox{Im} \; \psi_{5}(s)|_{HAD} =
2 f_{K}^{2} M_{K}^{4} \;\delta (s - M_{K}^{2}) +
\frac{1}{\pi} \; \mbox{Im} \; \psi_{5}(s)|_{K \pi \pi} 
\frac{[BW_{1}(s) + \lambda BW_{2}(s)]}{(1 + \lambda)}
\end{eqnarray*}
\begin{equation}
+ \; \frac{1}{\pi} \; \mbox{Im} \; \psi_{5}(s)|_{QCD} \theta (s - s_{0})\; ,
\end{equation}

where $f_{K} \simeq 1.2 f_{\pi}$, $\mbox{Im} \; \psi_{5}(s)|_{QCD}$ is
the perturbative QCD spectral function modelling the continuum which
starts at some threshold $s_{0}$, $BW_{1,2}(s)$ are Breit-Wigner forms
for the two kaon radial excitations, normalized to unity at threshold,
and $\lambda$ controls the relative importance of the second radial
excitation. The choice $\lambda \simeq 1$ results in a reasonable
(smaller) weight of the K(1830) relative to the K(1460).\\

We have solved the Laplace transform QCD sum rules using the values:
$<\alpha_{s} G^{2}> \simeq 0.024 \mbox{GeV}^{4}$
, $<\bar{s} s> \simeq <\bar{u} u> = -0.01
\mbox{GeV}^{3}$, and allowing $\Lambda_{QCD}$ and $s_{0}$ to vary in the
range: $\Lambda_{QCD} = 280 - 380 \mbox{MeV}$, and $s_{0}= 4 - 8 
\mbox{GeV}^{2}$. 
The results for $m_{s}(1 \mbox{GeV}^{2})$ are very stable against variations
in the Laplace variable $M^{2}$ over the wide range: 
$M = 1 - 4 \mbox{GeV}^{2}$.
In Figures 1 and 2 we show some typical results for $m_{s}(1 \mbox{GeV}^{2})$
as a function of $M^{2}$ for, respectively, 
$\Lambda_{QCD} = 280 \mbox{MeV}$, $s_{0}= 4  - 6 \mbox{GeV}^{2}$, and
$\Lambda_{QCD} = 380 \mbox{MeV}$, $s_{0}= 6 - 8 \mbox{GeV}^{2}$. 
Combining all results gives
\begin{equation}
{\bar m}_{s}(1 \mbox{GeV}^{2}) = 155 \pm 25 \mbox{MeV} \;.
\end{equation}

This result is consistent with the other determinations in the scalar
channel, Eqs.(6)-(8). The error given above originates exclusively 
from changes in the relevant parameters, and does not reflect possible
systematic uncertainties from the hadronic sector. 

\newpage
\begin{center}
{\bf Figure Captions}
\end{center}
Figure 1. The running strange quark mass $m_{s}(M^{2} = 1 \;\mbox{GeV}^{2})$
as a function of the Laplace variable $M^{2}$, for
$\Lambda_{QCD} = 280 \;\mbox{MeV}$. Upper and lower curves determine
the range obtained by varying $s_{0}$ in the interval:
$s_{0} = 4.0 - 6.0 \;\mbox{GeV}^{2}$.\\

Figure 2. The running strange quark mass $m_{s}(M^{2} = 1 \mbox{GeV}^{2})$
as a function of the Laplace variable $M^{2}$, for
$\Lambda_{QCD} = 380 \;\mbox{MeV}$. Upper and lower curves determine
the range obtained by varying $s_{0}$ in the interval:
$s_{0} = 6.0 - 8.0 \;\mbox{GeV}^{2}$.

  \end{document}